# Efficient conversion of anti-phase spin order of protons into $^{15}$N magnetization using SLIC-SABRE

Stephan Knecht,[a,b] Alexey S. Kiryutin,[c,d] Alexandra V. Yurkovskaya,[c,d] Konstantin L. Ivanov*,[c,d]

[a] *Dept. of Radiology, Medical Physics, Medical Center – University of Freiburg, Faculty of Medicine, University of Freiburg, Germany*

[b] *Molecular Imaging North Competence Center (MOIN CC), Section for Biomedical Imaging, Dept. for Radiology and Neuroradiology, University Medical Center Schleswig Holstein (UKSH), University of Kiel*

[c] *International Tomography Center, Siberian Branch of the Russian Academy of Science, Novosibirsk, 630090, Russia*

[d] *Novosibirsk State University, Novosibirsk, 630090, Russia*

* Corresponding author, e-mail: ivanov@tomo.nsc.ru

## Abstract

SABRE (Signal Amplification By Reversible Exchange) is a technique for enhancement of NMR (Nuclear Magnetic Resonance) signals, which utilizes parahydrogen (*p*H$_2$, the H$_2$ molecule in its nuclear spin state) as a source of non-thermal spin order. In SABRE experiments, *p*H$_2$ binds to an organometallic complex with a to-be-polarized substrate; subsequently, spin order transfer takes place and the substrate acquires non-thermal spin polarization resulting in strong NMR signal enhancement. In this work we argue that the spin order of H$_2$ in SABRE experiments performed at high magnetic fields is not necessarily the singlet order but rather anti-phase polarization, $S_{1z}S_{2z}$. Although SABRE exploits *p*H$_2$, i.e., the starting spin order of H$_2$ is supposed to be the singlet order, in solution S-T$_0$ conversion becomes efficient once *p*H$_2$ binds to a complex. Such a variation of the spin order, which becomes $S_{1z}S_{2z}$, has an important consequence: NMR methods used for transferring SABRE polarization need to be modified. Here we demonstrate that methods proposed for the initial singlet order may not work for the $S_{1z}S_{2z}$ order; however, a simple modification makes them efficient again. A theoretical treatment of the problem under consideration is proposed; these considerations are supported by high-field SABRE experiments. Hence, for efficient use of SABRE one should note that polarization formation is a complex multi-stage process: careful optimization of this process may not only deal with chemical aspects but also with the spin dynamics, including the spin dynamics of $H_2$.

# I. Introduction

SABRE (Signal Amplification By Reversible Exchange) [1] is a fast developing, parahydrogen ($pH_2$, the $H_2$ molecule in its nuclear spin state) based hyperpolarization technique for NMR (Nuclear Magnetic Resonance) signal enhancement. Polarization is generated within an organometallic complex where the to-be-polarized substrate and $pH_2$ form a coupled spin system, see **Scheme 1**. The substrate acquires polarization via spin order transfer from $pH_2$: such a transfer can be achieved at low [1-3] or ultralow [4-7] fields via coherent spin mixing and at high-fields via cross-relaxation [8, 9] or by using special radiofrequency (RF) excitation transfer schemes [10-15]. Spontaneous spin order transfer, i.e., transfer in the absence of NMR excitation, is efficient at low fields, ranging from mT for protons down to nT for X-nuclei. This however means that hyperpolarization needs to be generated outside of the NMR magnet (or an MRI scanner), which imposes the additional obstacle of operating external polarizers and consequently necessitating the transfer of the sample to high fields for measurements. Several approaches have been suggested recently to enable direct, in-bore, hyperpolarization which eliminates this shortcoming. Many high-field transfer schemes, however, fall short of their predicted efficiency [11, 12, 16]. In this work we show experimentally and theoretically that, at least for the substrate and solvent employed here (being a relatively standard SABRE chemical system), this is in part because of singlet-triplet mixing in $H_2$, which is explained below. Such a spin mixing results in a different initial spin order of $H_2$, which is the anti-phase spin order, $\hat{S}_{1z}\hat{S}_{2z}$. Evidence for formation of anti-phase spin order has been presented in previous works [9, 17]. Once the initial state of the hyperpolarized spin system has changed, in principle, existing polarization transfer schemes need to be revisited and, when necessary, updated to guarantee maximal transfer efficiency. Here we present an extended analysis of one such transfer scheme and report a new method (termed SLIC-SABRE) for manipulating the $\hat{S}_{1z}\hat{S}_{2z}$ spin order present in the SABRE system. The source of the spin mixing is most likely the inequivalent proton position of bound dihydrogen in a SABRE complex. It is worth mentioning, that such a complex does not need to be the one visible in NMR experiments, but might be a fast exchanging intermediate, possibly present in low concentration [17].

## II. Methods

### *A. High-field SABRE methods*

SABRE-derived polarization of a substrate is formed in the following way. Both $pH_2$ and the substrate reversibly bind to an Ir-based complex. Spin order transfer in the SABRE complex gives rise to polarized bound substrate, here pyridine (Py), in the equatorial positions. Exchange between the free and bound form of the substrate produces hyperpolarized free substrate molecules in solution, see **Scheme 1**. Hereafter the free and catalyst-bound forms of $H_2$ are termed $fH_2$ and Ir-$H_2$. Likewise, different forms of Py are termed fPy (free Py), ePy (equatorial Py) and aPy (axial Py).

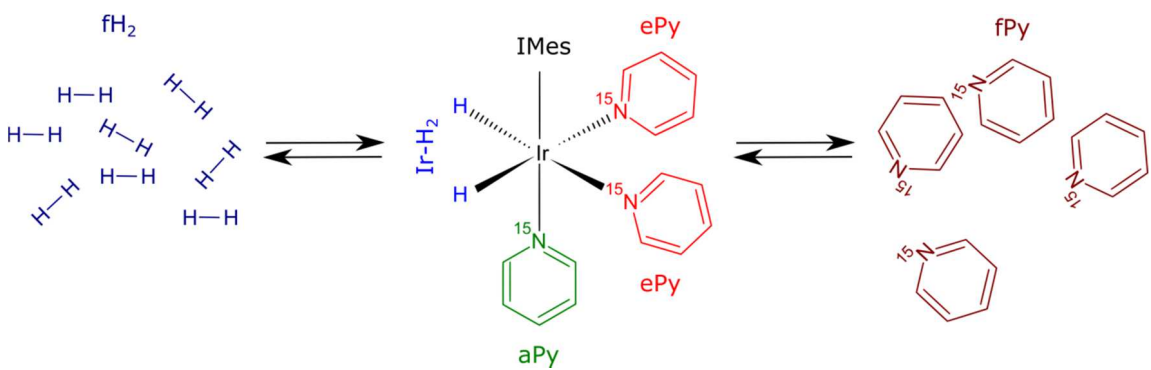

**Scheme 1**: Reaction kinetics of the SABRE system: Both free hydrogen in solution ($fH_2$) and the free substrate (Pyridine, fPy) exchange with the SABRE complex (here Ir-IMes complex); for the Py ligands exchange is most efficient for the equatorial positions. Spin order transfer occurs in the SABRE complex; it is facilitated inside the coupling network of the complex.

When a SABRE experiment is performed at a high external magnetic field polarization transfer can be due to cross-relaxation [8, 9]. Such a transfer pathway is known to be relatively inefficient; for this reason, NMR pulse schemes have been designed for efficient high-field SABRE polarization. Such schemes either exploit coherence transfer (they are based on the famous INEPT method) [14, 15] or spin mixing at Level Anti-Crossings (LACs) in the rotating frame. For the latter family of methods some authors use the term Spin Locking induced Crossing (SLIC) [18], meaning that spin order transfer is achieved by using continuous wave (CW) RF-fields, inducing energy matching (i.e., LAC conditions) in the rotating frame. The SLIC method can be exploited [12, 18, 19] for conversion between magnetization and singlet spin order in pairs of nearly equivalent spins. SLIC can also be applied to multi-spin systems [12] for the same purpose.

The first SLIC-based method to polarize $^{15}N$-nuclei by SABRE at high-fields has been introduced by Theis et al. [12]; the technique has been named LIGHT-SABRE. Assuming a small coupling difference between the two equatorial bound $^{15}N$ nuclei the RF-excitation scheme has been designed to utilize the (presumably small) magnetic non-equivalence of the Ir-H$_2$ protons to convert the singlet state into spin magnetization, in this case $x$-polarization, of the $^{15}$N spins of the ePy ligands. A subsequent selective pulse was used to generate $z$-magnetization of ePy, which would not dephase; exchange with the free substrate pool gives rise to fPy polarization, see **Figure 1**. Theoretical analysis predicts very high efficiency of polarization transfer, provided the state of Ir-H$_2$ is given by the singlet order, even when considering the real couplings of the SABRE complex (compare Table 1) which are far away from traditional SLIC conditions. More precisely, the spin order of Ir-H$_2$ we consider is given by the projection of the singlet state of H$_2$ onto the spin eigenstates of the SABRE complex, because the singlet state is no longer an eigenstate of the complex with H$_2$ and coherences will be "washed out" if they evolve faster than the exchange occurs [3, 20, 21]. At the same time, one should note that LIGHT-SABRE is not as efficient as expected from theoretical analysis, see examples given below. A related approach to high-field SABRE is given by a technique [10], which utilizes two RF-fields applied to protons and to $^{15}$N nuclei and spin mixing at an LAC in the doubly rotating frame. This method gives rise to signal enhancement, $\varepsilon$, of the order of 2,000 (about an order of magnitude higher than LIGHT-SABRE). Below, we discuss the reason for different signal enhancement provided by different methods.

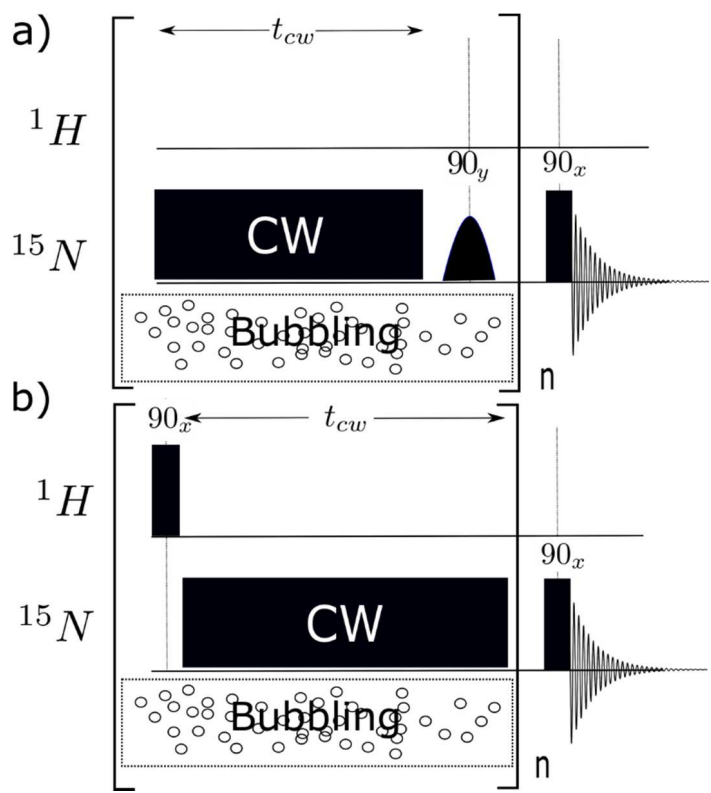

**Figure 1**: Top: Pulse scheme of LIGHT-SABRE suggested by Theis et al. to convert proton singlet spin order into $^{15}$N magnetization. The CW-field is applied on resonance with $^{15}$N-ePy for the time period of $t_{cw}$, followed by a selective soft pulse to convert transversal into longitudinal magnetization. Bottom: Pulse scheme, SLIC-SABRE, suggested in this work to convert the antiphase $\hat{S}_{1z}\hat{S}_{2z}$ proton spin order into longitudinal $^{15}$N magnetization. The 90 degree pulse on the $^1$H channel generates the $\hat{S}_{1z}\hat{S}_{2z}$ spin order which is then converted into longitudinal substrate magnetization by an off-resonant CW-field. Both methods are repeated multiple, $n$, times to increase polarization in the free substrate pool.

## B. Experimental protocols

In this work we have run LIGHT-SABRE experiments as described before [12]. Additionally, we propose and investigate a related but distinctly different SLIC-based scheme (which we term SLIC-SABRE), see **Figure 1**. Firstly, we chose to use an off-resonant instead of an on-resonant CW-field. This allows the direct generation of z-magnetization, eliminating the need for a selective soft pulse and thereby somewhat reducing the complexity of the experimental procedure. In our experiments this simplification provided comparable efficiency with respect to the utilization of a selective soft-pulse, as was done in the original implementation of LIGHT-SABRE (see below). We choose it not for the purposes of optimizing the efficiency, but because it makes experiments easier to implement and run as only one NMR pulse on the $^{15}$N channel is used. Secondly, considering the case where spin order of H$_2$ is given by $\hat{S}_{1z}\hat{S}_{2z}$ rather than the singlet order, we applied a 90-degree pulse prior to applying the CW-field on the $^{15}$N channel. In this manner, we can generate spin order suitable for manipulation by the subsequent CW-field. As in the original description of LIGHT-SABRE we repeatedly apply both methods during continuous bubbling of pH$_2$ to maximize the free substrate polarization.

All experiments were performed using CD$_3$OD as the solvent and $^{15}$N-Py as the substrate. Our pre-catalyst was IrCl(COD)(IMes) [22] (IMes = 1,3-bis(2,4,6-trimethylphenyl), COD = cyclooctadiene); hereafter Ir-IMes. All experiments were carried out with 90% enrichment of pH$_2$ under 2 bars of pH$_2$ pressure. The hydrogen flow rate was set to a low value, which ensured that most of the sample remained inside the NMR coil volume.

## III. Results

### A. Theory

*Spin order of H$_2$ in high-field SABRE experiments*

Most polarization transfer schemes in SABRE have been designed assuming that the starting spin order of the pair of protons (originating from pH$_2$) is the singlet spin order. Hence, these schemes attempt to maximize the conversion efficiency from the $|S\rangle$ state population of the protons into large net polarization of the substrate molecule. At first glance, this assumption is completely reasonable, given that H$_2$ in solution is rapidly replaced by externally supplied fresh pH$_2$. However, we recently found [9, 17] that there is a fast and efficient mixing between the singlet state and the central triplet state, $|T_0\rangle$, in catalyst-bound H$_2$. Such a behavior needs to be considered because it changes the spin order of H$_2$ in solution and thus the efficiency and behavior of the many polarization schemes. The $S$-$T_0$ mixing changes the spin order of H$_2$ resulting in the formation of the anti-phase spin order, $\hat{S}_{1z}\hat{S}_{2z}$. Detailed evidence for the formation of such a spin order in SABRE systems is presented in our previous work [9].

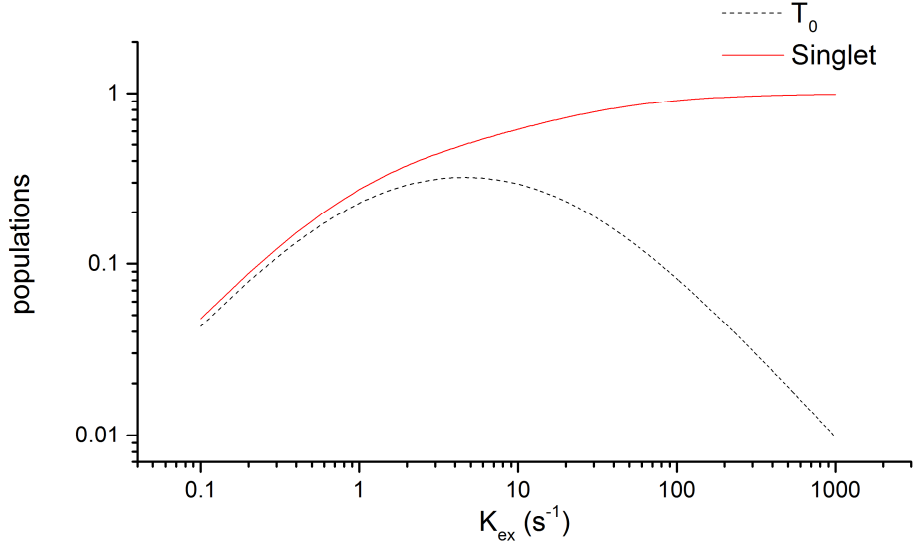

**Figure 2**: Populations of singlet ($|S\rangle$, red solid line) and $|T_0\rangle$ states (black dashed line) for different external exchange rates of $pH_2$. It is noteworthy, that there are two regimes: when mixing and relaxation are fast, compared to the externally supplied $pH_2$ there is almost no overpopulation of singlet state with respect to $|T_0\rangle$ and the resulting spin order is mainly $\hat{S}_{1z}\hat{S}_{2z}$, see eq. (2). When $pH_2$ is supplied significantly faster than the mixing and relaxation rates, the resulting state of $pH_2$ is the singlet spin order. Parameters used for simulations are $R_S = 1$ s$^{-1}$, $R_{T_0} = 1$ s$^{-1}$ and $\sigma = 10$ s$^{-1}$.

We suggest the following kinetic scheme for qualitative modeling of the spin order of H$_2$, which is present in solution and used as a source of NMR signal enhancement. By "H$_2$ is solution" we mean H$_2$, which is present in the free form of dissolved molecular hydrogen and then exchanges with the various catalyst-bound forms. When the two protons of catalyst-bound H$_2$ are chemically non-equivalent, i.e. they have different chemical shifts, coherent $S$-$T_0$ mixing becomes fast and efficient [23, 24]. The populations of the two relevant states of H$_2$ are denoted as $P_S$ (singlet-state population) and $P_{T_0}$ (central triplet state population). Below, instead of the state populations we introduce their deviation from the corresponding values at thermal equilibrium, denoted as $\delta P_S$ and $\delta P_{T_0}$. Both $\delta P_S$ and $\delta P_{T_0}$ relax to zero with their respective rates $R_S$ and $R_{T_0}$. In addition the transition rate $\sigma$ between those states describes the effects of singlet-triplet mixing. The whole pool is under constant exchange with the externally supplied $pH_2$ at a rate $K_{ex}$. This rate is introduced purely empirically and will depend strongly on the specific setup and the experimental conditions. The above considerations lead to the following differential equation for the spin-state of solution H$_2$:

$$\frac{d}{dt}\begin{pmatrix} \delta P_S \\ \delta P_{T_0} \\ P_S^{ex} \end{pmatrix} = \begin{pmatrix} -R_S - \sigma - K_{ex} & \sigma & K_{ex} \\ \sigma & -R_{T_0} - \sigma - K_{ex} & 0 \\ 0 & 0 & 0 \end{pmatrix} \begin{pmatrix} \delta P_S \\ \delta P_{T_0} \\ P_S^{ex} \end{pmatrix} \quad (1)$$

Here $P_S^{ex}$ stands for amount of the singlet order provided by the external supply. The plot in **Figure 2** shows the dependence of steady-state $|S\rangle$ and $|T_0\rangle$ populations on the external supply rate of $p$H$_2$. When the latter is low compared to relaxation and singlet-triplet mixing rate, $\sigma$, the resulting $|S\rangle$ and $|T_0\rangle$ populations are almost equal (and relaxation limited). Only when the external supply outruns the relaxation processes and $S$-$T_0$ mixing, the spin state of H$_2$ becomes the singlet state. The experiments we report below suggest that we are in the regime of almost perfect mixing. Potentially spin mixing can be reduced by initiating the SABRE process faster and applying the NMR pulse sequence for polarization transfer before the steady-state is reached for the spin order of H$_2$. This is in accordance with previous reports of strongly increasing polarization for faster bubbling of $p$H$_2$.

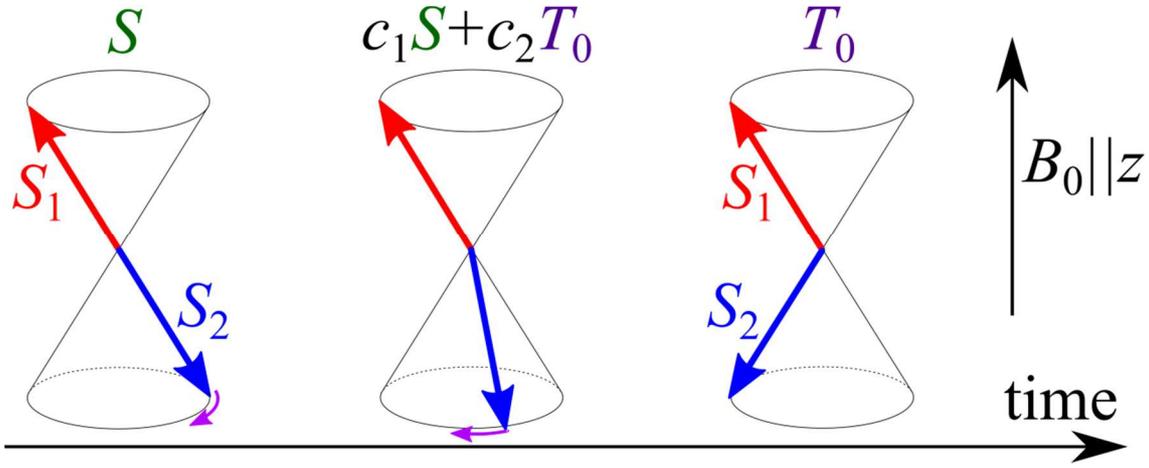

**Scheme 2**: Vector model explaining the mechanism of the $S$-$T_0$ conversion. Here the red and blue arrows show the spin vectors of the two spins; at high $B_0$ field each spin precesses on a cone about the $z$-axis, here $z||B_0$. The singlet state $|S\rangle$ is the state with anti-parallel spins. In the $|T_0\rangle$ state there is no net $z$-magnetization, but the total spin is non-zero. $S$-$T_0$ transitions in a spin pair occur due to the difference, $\delta\nu$, in the NMR precession frequency of the two spins. Consequently, a spin pair starting from the singlet state transforms to a superposition of the $S$ and $T_0$ states, then to $T_0$ and back.

*Spin mixing pathways*

Let us now elucidate, which polarization transfer pathways become active in different transfer schemes. We investigated the transfer efficiency for different initial spin orders of H$_2$ in solution given by the density matrix $\rho_0$. Here we restrict ourselves to two limiting cases:

$$\rho_0 = \frac{1}{2}(|S\rangle\langle S| + |T_0\rangle\langle T_0|) = \frac{1}{4}\hat{E} - \hat{S}_{1z}\hat{S}_{2z}$$

$$\rho_0 = |S\rangle\langle S| = \frac{1}{4}\hat{E} - (\hat{\mathbf{S}}_1 \cdot \hat{\mathbf{S}}_2) = \frac{1}{4}\hat{E} - \hat{S}_{1z}\hat{S}_{2z} - \hat{S}_{1x}\hat{S}_{2x} - \hat{S}_{1y}\hat{S}_{2y}$$

(2)

In the latter matrix the $\hat{S}_{1x}\hat{S}_{2x}$ and $\hat{S}_{1y}\hat{S}_{2y}$ correspond to the Zero-Quantum Coherence (ZQC) [25]: the $S$-$T_0$ mixing is driven by the evolution of this coherence as explained in **Scheme 2** using a well-known vector model, which is widely used in spin chemistry literature [26-34]. The ZQC evolution is operative when the two spins are weakly coupled, that is, the difference, $\delta\nu$, in their Zeeman interaction with the external magnetic fields is much greater than the coupling between them, $J_{HH}$. When $pH_2$ is attached to different complexes at different instants of times the ZQC has a different phase in different complexes. Consequently, the coherence is washed out at the density matrix of the spin system is given by the $\hat{S}_{1z}\hat{S}_{2z}$ term; in this situation the populations of the $|S\rangle$ and $|T_0\rangle$ states exactly coincide.

Now let us describe how polarization transfer to the nitrogen spins occurs. For simplicity, we consider the situation where RF-fields are applied only to nitrogen spins, as it is done, e.g., in LIGHT-SABRE. The proton spin operators are denoted as $\hat{\mathbf{S}}_{1,2}$, the nitrogens are $\hat{\mathbf{I}}_{1,2}$. We always consider a four-spin system, two protons coupled to two nitrogen spins, of an AA′XX′ kind. Hence, the chemical shifts of spins are pair-wise equal. Likewise, there are two pairs of identical proton-nitrogen couplings, $J_{NH} = J_{N'H'}$ and $J_{NH'} = J_{N'H}$. The Hamiltonian of the spin system in the rotating frame is as follows (the frame rotation is done only for the $^{15}$N spins) [12]:

$$\hat{\mathcal{H}}^{rf} = -\nu_H(\hat{S}_{1z} + \hat{S}_{2z}) - \nu_1(\hat{I}_{1x} + \hat{I}_{2x}) + J_{HH}(\hat{\mathbf{S}}_1 \cdot \hat{\mathbf{S}}_2) + J_{NN}(\hat{\mathbf{I}}_1 \cdot \hat{\mathbf{I}}_2) + \hat{V} \tag{3}$$

Here we assume resonant RF-excitation (the RF-frequency $\nu_{rf}$ matches the $^{15}$N NMR frequency $\nu_N$) with $\nu_1$ is the RF-field strength; $\nu_H$ stands for the proton NMR frequency, $J_{HH}$ and $J_{NN}$ are the HH and NN spin-spin couplings, respectively. The perturbation term comes from the HN-couplings. It is as follows (for simplicity, we set two couplings to zero, $J_{NH'} = J_{N'H} = 0$):

$$\hat{V} = J_{NH}(\hat{S}_{1z}\hat{I}_{1z} + \hat{S}_{2z}\hat{I}_{2z}) \tag{4}$$

Now let us tilt the reference frame for the N-spins: $x \to z$ and $z \to -x$. The Hamiltonian in this "rotating tilted" frame becomes:

$$\hat{\mathcal{H}}^{tf} = -\nu_H(\hat{S}_{1z} + \hat{S}_{2z}) - \nu_1(\hat{I}_{1z} + \hat{I}_{2z}) + J_{HH}(\hat{\mathbf{S}}_1 \cdot \hat{\mathbf{S}}_2) + J_{NN}(\hat{\mathbf{I}}_1 \cdot \hat{\mathbf{I}}_2) + \hat{V}' \tag{5}$$

Here the frame is tilted for the N-spins and not tilted for the H-spins. The re-defined perturbation term changes to:

$$\hat{V}' = -J_{NH}(\hat{S}_{1z}\hat{I}_{1x} + \hat{S}_{2z}\hat{I}_{2x}) \tag{6}$$

The $\hat{V}'$ term gives rise to the following spin dynamics. It can drive the $S \to T_0$ transitions for the protons and spin-flipping transitions for N-spins, enriching or depleting the $|T_\pm\rangle$ states. Hence, the N-spins can be net-polarized along the effective field, in this case along the $x$-axis of the non-tilted frame. To convert polarization into longitudinal spin order, in LIGHT-SABRE an additional 90-degree pulse is applied [12]. The relevant blocks of the spin Hamiltonian where spin mixing occurs have been specified before [12]. Let us consider just one of these blocks, spanned by the states

$$|SS\rangle, \quad |T_0 T_0\rangle, \quad |T_0 T_+\rangle, \quad |T_0 T_-\rangle \tag{7}$$

Hereafter, in the ket-notations of the states of the four-spin system under consideration, the first symbol denotes the spin state of the two protons and the second symbols stands for the spin states of the nitrogens. If we neglect the $\hat{V}'$ term the energies of these spin states are:

$$E(SS) = -\frac{3}{4}J_{HH} - \frac{3}{4}J_{NN}, \quad E(T_0 T_0) = \frac{1}{4}J_{HH} + \frac{1}{4}J_{NN},$$

$$E(T_0 T_+) = \frac{1}{4}J_{HH} + \frac{1}{4}J_{NN} - \nu_1, \quad E(T_0 T_-) = \frac{1}{4}J_{HH} + \frac{1}{4}J_{NN} + \nu_1 \tag{8}$$

The perturbation term can cause mixing of the kind $SS \leftrightarrow T_0 T_+$ and $SS \leftrightarrow T_0 T_-$ because the perturbation drives the $S \to T_0$ transitions for the protons and flips the N-spins driving the $S \to T_\pm$ transitions. Relevant spin mixing occurs when the LAC conditions are fulfilled. For the first LAC this happens when

$$E(SS) = E(T_0 T_+) \Rightarrow J_{HH} + J_{NN} = \nu_1 \tag{9}$$

and for the second LAC this happens when

$$E(SS) = E(T_0 T_-) \Rightarrow J_{HH} + J_{NN} = -\nu_1 \tag{10}$$

Hence, the generalized matching condition is as follows:

$$\nu_1 = \pm(J_{HH} + J_{NN}) \tag{11}$$

For the other block of the Hamiltonian the matching conditions are

$$\nu_1 = \pm(J_{HH} - J_{NN}) \tag{12}$$

One should note that the present analysis is also valid when an off-resonant RF-field is applied on the $^{15}$N-channel. In this situation, in eqs. (11) and (12) one should replace $\nu_1$ by the strength of the effective field, $\nu_{eff}$, experienced by the $^{15}$N-spins in the rotating frame. The $^{15}$N-spins will be polarized along the effective field: hence, when this field has a z-component longitudinal polarization of the $^{15}$N-spins will be generated even without an additional RF-pulse.

After spin mixing the initial population difference of the proton $|S\rangle$ and $|T_0\rangle$ states, $\delta P_H$, is converted (in the ideal case, completely) into the population difference of the nitrogen $|T_\pm\rangle$ and $|S\rangle$ states. For different spin order, given by the proton spin density matrix $\rho_H^0$, the $\delta P_H$ value is as follows:

$$\begin{cases} \rho_0 = \frac{1}{4}\hat{E} - (\hat{\mathbf{S}}_1 \cdot \hat{\mathbf{S}}_2) \Rightarrow \delta P_H = 1 \\ \rho_0 = \frac{1}{4}\hat{E} - \hat{S}_{1z}\hat{S}_{2z} \Rightarrow \delta P_H = 0 \\ \rho_0 = \frac{1}{4}\hat{E} - \hat{S}_{1x}\hat{S}_{2x} \Rightarrow \delta P_H = 1/2 \end{cases} \tag{13}$$

Hence, for the singlet spin order the LIGHT-SABRE pulse sequence should work perfectly. For the $\hat{S}_{1z}\hat{S}_{2z}$ order LIGHT-SABRE should not work at all: there is simply no population difference of the proton $|S\rangle$ and $|T_0\rangle$ states. There are two ways to tackle this problem. One possible option is to apply a 90-degree pulse to perform spin order conversion $\hat{S}_{1z}\hat{S}_{2z} \to \hat{S}_{1x}\hat{S}_{2x}$: in this situation the desired population difference of the proton $|S\rangle$ and $|T_0\rangle$ states is recovered. Another option is to exploit the population difference between the proton $|S\rangle$ and $|T_\pm\rangle$ states. This can be done, for instance, by applying RF-fields on both channels as has been done before [10]. In this case one can mix, for instance, $|SS\rangle$ and $|T_+T_-\rangle$. There are more LACs in the system (altogether there are 8 LACs), which can lead to spin mixing of the desired kind.

In both cases there will be a decrease in the efficiency compared to the ideal case of the starting proton singlet order. If we start from $\hat{S}_{1x}\hat{S}_{2x}$ we lose by a factor of two (this cannot be overcome because the loss of spin order given by ZQC is irreversible).

Hence, the above analysis provides experimental strategies to modify the LIGHT-SABRE technique to make it efficient when the starting proton spin order is the anti-phase spin order. We simulated numerically the efficiency of the different transfer schemes shown in **Figure 1**. These simulations are based on our recently formulated [35] general theory of SABRE (a detailed description will be provided elsewhere). We explicitly consider the spin dynamics in the SABRE complex as well as chemical exchange between the free and catalyst-bound forms of the substrate (defining the complex dissociation rate $k_d$). Furthermore, we take into account the NMR parameters of both the complex and the free substrate, namely: relaxation rates, J-couplings and the effect of the CW-field in the rotating frame of reference. The spin state of $H_2$ is considered in the extreme cases of being either the singlet state or $\hat{S}_{1z}\hat{S}_{2z}$. The parameters for the simulations are reported in the caption of **Figure 3** and in Table 1.

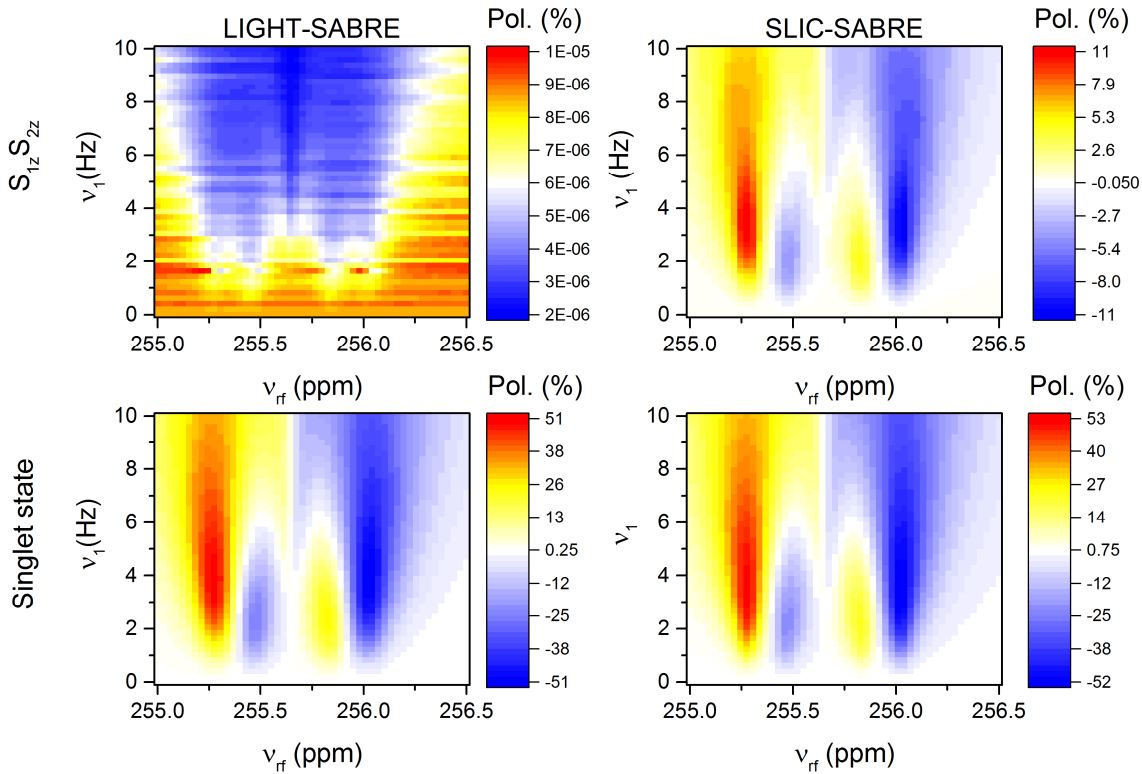

**Figure 3**: Theoretical dependence of the steady-state fS polarization on the frequency and amplitude of the RF-field. Here we compare the performance of LIGHT-SABRE (left) and SLIC-SABRE (right) for the two different cases of $H_2$ spin-order, namely $\hat{S}_{1z}\hat{S}_{2z}$ (top) and singlet-state (bottom). Simulations assume a four-spin system and coupling constants as listed in Table 1. The simulation parameters are: $k_d = 10$ s$^{-1}$, $\frac{[fS]}{[C]} = 30$; spin relaxation rates are $\frac{1}{30}$ s$^{-1}$ (for free substrate), $\frac{1}{3}$ s$^{-1}$ (for bound substrate) and 1 s$^{-1}$ (for Ir-H$_2$). Polarization is shown by contour maps and it is given in percent.

From the results of these simulations (see **Figure 3**) it becomes clear that LIGHT-SABRE is effective only when the spin state of $H_2$ is the singlet state, as expected from the theoretical analysis presented above. The SLIC-based sequence we suggest in this work, however, works for both spin orders, with one significant detail: when the initial spin order is the singlet order, the 90-degree pulse is no longer needed and z-magnetization is generated directly by the off-resonant CW-field. When the spin order is $\hat{S}_{1z}\hat{S}_{2z}$, the 90-degree pulse in the beginning is needed to generate the appropriate initial spin order, which evolves into magnetization of the substrate. It should be noted, that we completely neglected relaxation of the $H_2$ spin order;

reducing the steady-state populations. In combination with experimental imperfections, such as motion of molecules in and out of the coil volume, line-broadening coming from the gas bubbles in the NMR tube; our experiments fall short of the predicted efficiency (compare experimental results).

**Table 1**. Simulation parameters used in Figure 3.

| $J$ (Hz) | $H_1$ | $H_2$ | $^{15}N_1$ | $^{15}N_2$ |
|---|---|---|---|---|
| $H_1$ |  | −7.7 | −20.91 | 0.63 |
| $H_2$ |  |  | 0.63 | −20.91 |
| $^{15}N_1$ |  |  |  | 0.39 |
| $^{15}N_2$ |  |  |  |  |
| $\delta$ (ppm) | −22.6 | −22.6 | 255.3 | 255.3 |

## B. Experimental results

### SLIC-SABRE Experiments

We implemented and carried out the experiments according to **Figure 1**. These results are presented in **Figure 4**. The original LIGHT-SABRE method produces results comparable to SLIC-SABRE method without the initial 90 degree proton pulse. In both cases the line is distorted by two-spin order involving $^1$H and $^{15}$N nuclei. Both methods however show enhancements more than an order of magnitude lower than the scheme employing a 90-degree pulse on the hydrogen channel. Hence, the additional $^1$H-NMR pulse recovers the desired singlet spin order rendering the transfer scheme efficient. The pulsed SLIC-SABRE method we propose here turns out to be highly efficient, even under the experimental difficulties coming from gas bubbling. The drastic change in efficiency of the transfer schemes with and without the additional pulse allows us to conclude (as supported by the simulations carried out above and previous studies [9, 17]) that the dominant spin order of H$_2$ is not the singlet spin order, but the anti-phase spin order, $\hat{S}_{1z}\hat{S}_{2z}$.

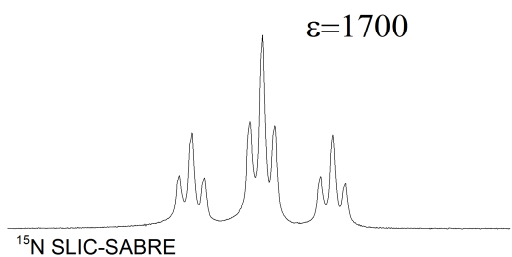

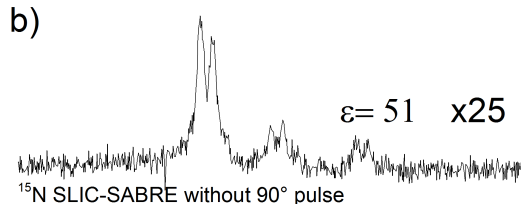

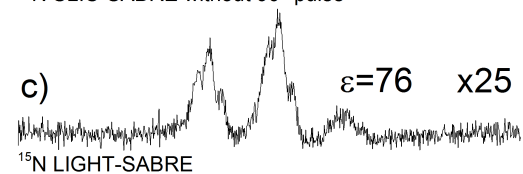

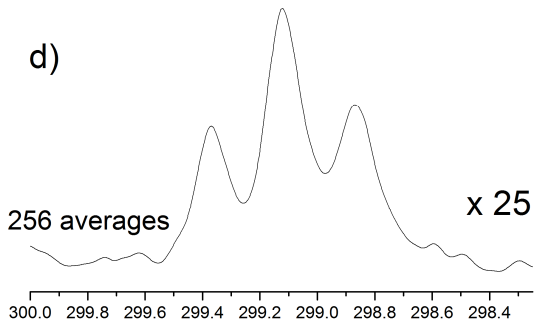

Figure 4: a) $^{15}$N NMR spectrum acquired after 300 repetitions (total experimental time of 60 s) of the SLIC-SABRE sequence proposed in this work. The CW-field frequency was set to 255.3 ppm with an amplitude of 5 Hz and $t_{cw} = 0.2$ s. b) $^{15}$N NMR spectrum acquired after 600 repetitions of the same pulse scheme as in a) but without the 90-degree pulse in the beginning. c) $^{15}$N NMR spectrum acquired after 300 repetitions of the LIGHT-SABRE scheme ($t_{cw} = 0.2s$, RF amplitude is 10 Hz. d) Thermal NMR spectrum after 256 averages (repetition time 120 seconds). In the spectrum we applied line-broadening by 3 Hz. Concentrations are: 1 mM of the Ir-IMes catalyst, 40 mM fPy. All hyperpolarized spectra were acquired using a continuous bubbling setup with low flow rates at 2 Bar pressure using 90% enriched $pH_2$. For the SABRE spectra signal enhancement factors, $\varepsilon$, are specified.

*Comparison with SABRE-INEPT*

In order to complete the picture, we furthermore compare the SLIC-based scheme to our recent adaptation [36] of SABRE-INEPT, see **Figure 5**. One can clearly see that multiple repetition of the pulse sequence (using selective pulses for polarization transfer) allows one to improve strongly the resulting enhancement. At the same time, SABRE-INEPT (which is expected to work well for both anti-phase spin order and for the pure singlet order) has better performance than the original LIGHT-SABRE without modification. Hence, one can see that at our experimental conditions polarization transfer is efficient when (i) the transfer scheme is properly modified to generate the singlet spin order or (ii) it is operative for both singlet spin order and anti-phase spin order.

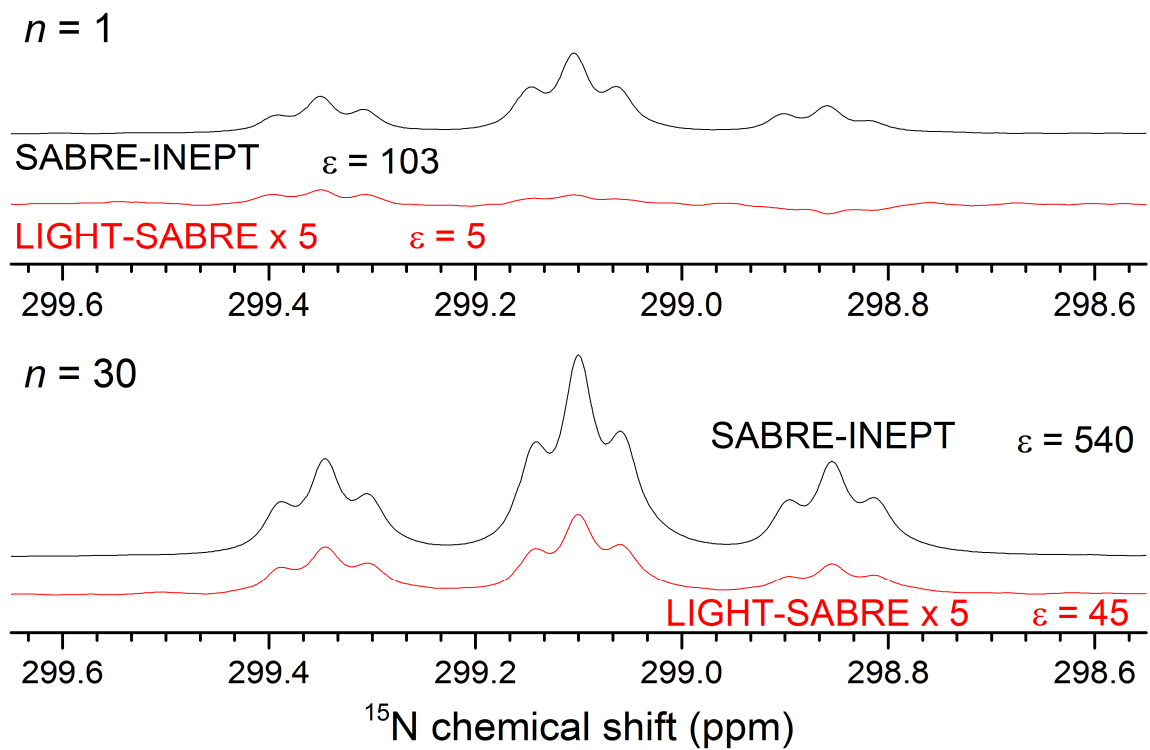

**Figure 5**. Comparison of the performance of LIGHT-SABRE and SABRE-INEPT for $n = 1$ (top, single application of the pulse sequence) and $n = 30$ (bottom, re-polarization). Concentrations are: 2 mM of the Ir-IMes catalyst and $[C_{fS}] = 44$ mM. Sample temperature was 12.5 °C and at 2.5 bar $pH_2$ pressure; $pH_2$ enrichment was 50%.

## IV. Summary and conclusions

In this work we argue that the spin order of $H_2$ in high-field SABRE experiments is often the anti-phases spin order rather than singlet order. This finding has important consequences for the spin dynamics in high-field SABRE; furthermore, modification of the spin order can lead to reduced spin order transfer efficiency and requires updating transfer schemes used in high-field SABRE. Here we report a new method to convert $\hat{S}_{1z}\hat{S}_{2z}$ spin order into $^{15}N$ magnetization in an efficient way using SLIC-SABRE. We give a qualitative understanding of the spin mixing and its dependence on relaxation parameters and the external supply rate of $pH_2$. From this data we conclude that there is a regime, in which $\hat{S}_{1z}\hat{S}_{2z}$ spin order dominates, and one, in which singlet order does. Our theoretical examination of the performance of the transfer schemes used in this work in combination with the experimental results shows clearly that we are operating in the first of the two regimes and there is only a little overpopulation of the singlet state. Our study is likely to stimulate further investigations in this research direction, especially in light of the very recent ADAPT-SABRE method of Stevanato and coauthors [37, 38], where they introduced and successfully implemented a method transferring overpopulations of the singlet state (more precisely singlet-state of $H_2$ in solution). Specifically, it will be interesting to see if the difference in spin order of the two experiments arises due to the different substrates used (i.e., differences in the SABRE chemistry) or has its origin in the effectiveness of the bubbling setups at different labs. In general, we expect that the effects discussed here are dependent on the sample, experimental conditions and setup. Furthermore, the process of singlet-triplet mixing needs to be further investigated and strategies to preserve the singlet state in solution should be explored. When there is little or no singlet-triplet mixing, we expect LIGHT-SABRE to be highly effective in its original formulation: modification discussed here is then no longer needed. Our study suggests that SABRE polarization is a complex process comprising additional stages, which were neglected in previous works but may play an important role in polarization formation.

## Acknowledgements

This work has been supported by the Russian Science Foundation (grant No. 14-13-01053); we are thankful to FASO of RF (project 0333-2017-0002) for providing access to NMR facilities. S.K. acknowledges DAAD for a fellowship for coming to ITC (Novosibirsk, Russia) and the Emmy Noether program of the DFG (HO 4604/2-1) for funding of his PhD studies when in Germany. We acknowledge Prof. Warren S. Warren (Duke University) for useful discussions.